\documentclass[aps,prb,twocolumn,superscriptaddress, show pacs]{revtex4-1}
\usepackage{bm, amsmath}
\usepackage{graphicx}
\usepackage{color}%font colors
\usepackage{epstopdf}
\usepackage{graphicx}% Include figure files
\usepackage{dcolumn}% Align table columns on decimal point
\usepackage{bm}% bold math
\usepackage{subfigure}

\begin{document}

\title{Asymmetric isolated skyrmions in polar magnets with easy-plane anisotropy}

%Racetrack memory design based on attractive skyrmions in the cone phase of chiral magnets

%\title{Isolated chiral skyrmions and their attractive interactions \\
%in the cone phase of cubic helimagnets}

\author{A.~O.~Leonov}
\thanks{leonov@hiroshima-u.ac.jp}
\affiliation{Center for Chiral Science, Hiroshima University, Higashi-Hiroshima, Hiroshima 739-8526, Japan}
\affiliation{Department of Chemistry, Faculty of Science, Hiroshima University Kagamiyama, Higashi Hiroshima, Hiroshima 739-8526, Japan}
%\affiliation{IFW Dresden, Postfach 270016, D-01171 Dresden, Germany}

\author{I. K\'ezsm\'arki}
\affiliation{Experimental Physics V, Center for Electronic Correlations and Magnetism, University
of Augsburg, Augsburg 86135, Germany}
\affiliation{Department of Physics, Budapest University of Technology and Economics and MTA-BME Lend\"ulet
Magneto-Optical Spectroscopy Research Group, Budapest 1111, Hungary}

\date{\today}

\begin{abstract}
{We introduce a new class of isolated magnetic skyrmions emerging
within tilted ferromagnetic phases of polar magnets with easy-plane
anisotropy.
The asymmetric magnetic structure of these skyrmions is associated
with an intricate pattern of the energy density, which exhibits
positive and negative asymptotics with respect to the surrounding
state with a ferromagnetic moment tilted away from the polar axis.
Correspondingly, the skyrmion-skyrmion interaction has an
anisotropic character and can be either attractive or repulsive
depending on the relative orientation of the skyrmion pair.
We investigate the stability of these novel asymmetric skyrmions
against the elliptical cone state and follow their transformation
into axisymmetric skyrmions, when the tilted ferromagnetic moment of
the host phase is reduced.
Our theory gives clear directions for experimental studies of
isolated asymmetric skyrmions and their clusters embedded in tilted
ferromagnetic phases.
%
%Skyrmions are topological spin textures of interest for fundamental science and applications.
%
%Previous studies usually focus on the axisymmetric skyrmions stabilized by easy-axis anisotropy.
%Our theory gives clear directions for experimental studies of isolated skyrmions within angular phase.
}
\end{abstract}

% %%% PACS numbers

\maketitle

%%%%%%%%%%%%%%%%%%%%%%%%%%%%%%%%%%%%%%%%%%%%%%%%%%%%%%%%%%%%%%%%%%%%%%%%%%%%%%%%%%%%%%%%%%%%%%%%%%%%%%%%%%%%%%%%
%1.Introduction
%%%%%%%%%%%%%%%%%%%%%%%%%%%%%%%%%%%%%%%%%%%%%%%%%%%%%%%%%%%%%%%%%%%%%%%%%%%%%%%%%%%%%%%%%%%%%%%%%%%%%%%%%%%%%%%

%definition of skyrmion
Magnetic chiral skyrmions are particle-like topological solitons
with complex spin structure
\cite{Bogdanov94,Bogdanov89,review,Nagaosa13} which are the
solutions of the field equations of the Dzyaloshinskii's theory
\cite{Dz64}.
%
%Magnetic skyrmions are particle-like topological solitons with complex spin structure \cite{Bogdanov94,Bogdanov89,review,Nagaosa13} which can emerge as the solutions of the field equations of the Dzyaloshinskii's theory \cite{Dz64}.
%
Recently, skyrmion lattice states and isolated skyrmions were
discovered in bulk crystals of chiral magnets near the magnetic
ordering temperatures \cite{Muehlbauer09,Wilhelm11,Kezsmarki15} and
in nanostructures with confined geometries over larger temperature
regions\cite{Yu10,Yu11,Du15,Liang15}.
The small size, topological protection and easy manipulation of
skyrmions by electric fields and currents
\cite{Schulz12,Jonietz10,Hsu17} generated enormous interest in their
applications in information storage and processing
\cite{Sampaio13,Tomasello14}.
Depending on the crystal symmetry of the host materials, distinct
classes of skyrmions, such as Bloch and  N\'eel skyrmions, or
anti-skyrmions \cite{Nayak17} can be realized.
In particular, N\'eel skyrmions were recently found in GaV$_4$S$_8$
and GaV$_4$Se$_8$, which are magnetic semiconductors with non-chiral
but polar crystal structure \cite{Kezsmarki15,Bordacs17}. N\'eel
skyrmions emerging in such multiferroic hosts are associated with an
electric polarization pattern, which can be exploited for their
electric field control \cite{Ruff15}.
\begin{figure}[t!]
\includegraphics[width=0.95\columnwidth]{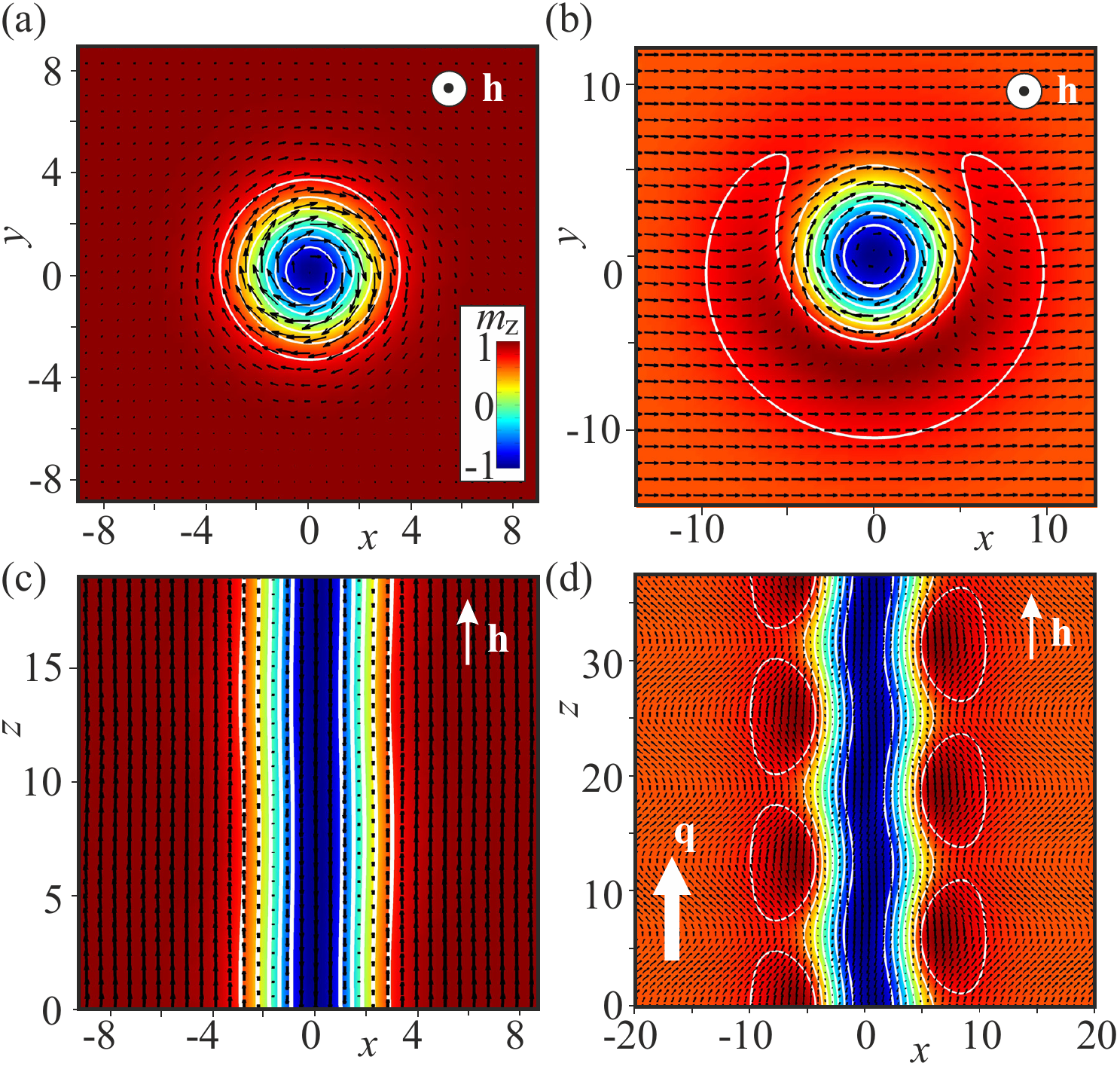}
\caption{ (color online) Cross-sections of the internal magnetic
structures of skyrmions obtained in model (\ref{density}) with DMI
(\ref{Bloch}) and no axial anisotropy, $k_u=0$. The magnetic field
$\mathbf{h}$ is directed along the $z$ axis. (a) \& (c) Color plots
of $m_z$ for an axisymmetric skyrmion embedded within the PFM state
($h$ = 0.55) in the $xy$ and $xz$ planes, respectively. (b) \& (d)
the same for a non-axisymmetric skyrmion in the conical phase ($h$
=0.3), with an additional screw-like rotation alongside with the
conical phase (white arrow shows the modulation vector $\mathbf{q}$
of the conical phase). The color bar in panel (a) is common for each
panel. The in-plane component of the magnetization is represented by
black arrows.}
%The color bar in panel(a) should have labeled as m_z instead of m.
%The black arrows representing the in-plane magnetization are not possible to see in panels (b) & (d). Can you do some coarse graining, i.e. only plot every 3rd or 4th spin but with larger arrows?
%Please put the arrows and the labels "h" into the top right corner for panels (c) & (d) as well. Then, in panel (c) you may move the "q" vector to the left side.
\label{FigBloch}
\end{figure}

The current interest of skyrmionics is focused on isolated \textit{axisymmetric} skyrmions %within the FM state saturated opposite to the magnetization in the skyrmion center.
within the polarized ferromagnetic (PFM) state of
non-centrosymmetric magnets. All the spins around such skyrmions are
parallel to the applied magnetic field and point opposite to the
spin in the center of the skyrmion, as visualized in Figs.
\ref{FigBloch} (a) \& (c).
The internal structure of such axisymmetric skyrmions, generally
characterized by repulsive skyrmion-skyrmion interaction, has been
thoroughly investigated theoretically \cite{Bogdanov94b,LeonovNJP16}
and experimentally by spin-polarized scanning tunneling microscopy
in PdFe bilayers with surface induced Dzyaloshinskii-Moriya
interactions (DMI) and strong easy-axis anisotropy
\cite{Romming13,Romming15}.
The existence region of axisymmetric skyrmions was found to be
restricted by strip-out instabilities at low fields and a collapse
at high fields \cite{LeonovNJP16}.

\begin{figure}
\includegraphics[width=0.95\columnwidth]{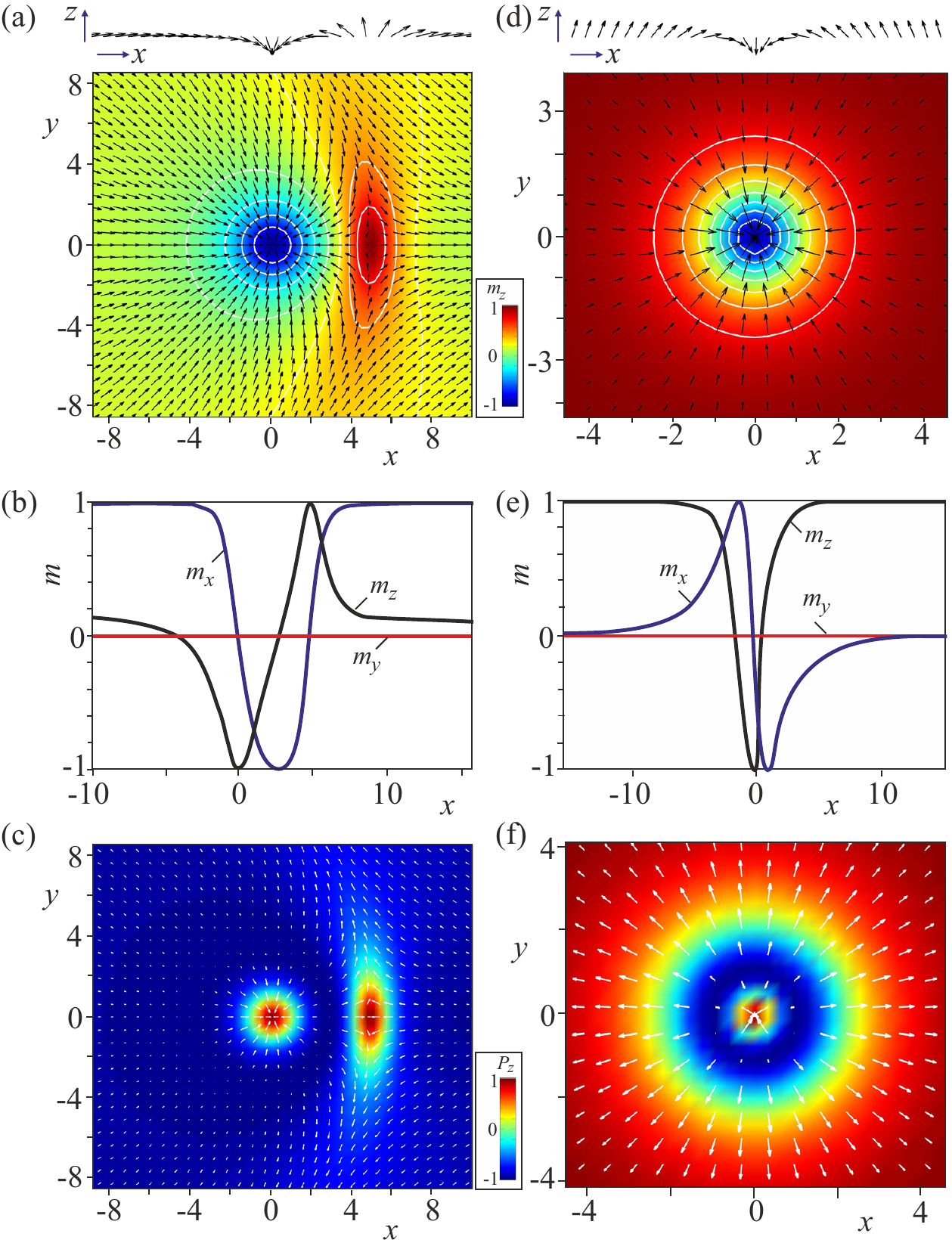}
\caption{ (color online) Numerical solutions for skyrmions using
model (\ref{density}) with DMI (\ref{DMI}). (a) - (c) Asymmetric
skyrmions within the TFM state for $h=0.5$ and $k_u=-1.1$. %corresponding to point $A$ in the phase diagram shown in Fig. \ref{PD} (a).
(d) - (f)
Axisymmetric skyrmions within the PFM state for $h=1.6$ and $k_u=-0.7$. %corresponding to point $B$ in the same phase diagram.
The top panels are color plots of $m_z$ in the $xy$ plane with black
arrows representing the in-plane magnetization components. 
The top insets show the magnetization components in the plane $xz$ across the skyrmion centers, which are additionally shown  as curves in the middle panels.  
%
%Curves in the middle panels show the magnetization components in the $xz$ cross-section through the centers of the skyrmions. 
Note that the $m_x$ and $m_z$ components seemingly exchange roles in panels (b) \&
(e), which is due to the different embedding phase, the TFM state
with nearly in-plane magnetization and the PFM state with fully
out-of-plane magnetization. %The same magnetization components across the skyrmion centers are also shown as black arrows in top insets in (a) and (d).
The bottom panels are color plots of
$P_z$ with white arrows representing the in-plane components of the
electric polarization, calculated according to Eq.
(\ref{polarization}).}
%It would help better visualizing the structure, if you could add a narrow panel showing the rotation of the spins along the x axis for y=0 for both the asymmetric and axis symmetric skyrmion, similarly to the schematic view in Fig. 1 of our NatMater paper on GaV4S8. I think it is a key point concerning our main message here.
%The polarization in panel (f) should be plotted over the same x-y range as the magnetization in panel (d).
%Labels (P_z, m_z) and numbers in the color bars should be horizontal.
%In panels (a) and (c) the arrows are hard to see. Is it possible to show only every second but with larger arrow?
\label{structure}
\end{figure}

The internal spin pattern of isolated skyrmions can break the
rotational symmetry once placed into the conical phase of bulk
helimagnets \cite{LeonovJPCM16}, such as the cubic B20 compounds.
These skyrmions are not uniform along their axes. Though their
central core region nearly preserves the axial symmetry, the
domain-wall region, which connects the core with the embedding
conical state, is \textit{asymmetric} \cite{LeonovJPCM16}. This
asymmetric profile of the cross-section forms a screw-like
modulation along the skyrmion core, as depicted in Figs.
\ref{FigBloch} (b) \& (d).
These asymmetric isolated skyrmions, which can exhibit an
\textit{attractive} skyrmion-skyrmion interaction, were proposed to
underlie the precursor phenomena near the ordering temperatures in
chiral B20 magnets (MnSi, FeGe) \cite{LeonovJPCM16} and have
prospects in spintronics as an alternative to the common
axisymmetric skyrmions\cite{LeonovAPL16}.

In this Letter we introduce a new type of isolated skyrmions within
the tilted ferromagnetic (TFM) state of magnets with polar crystal
structure and easy-plane anisotropy. Such skyrmions are forced to
develop an asymmetric shape in order to match their spin pattern
with that of the TFM state, meanwhile preserving their topological
charge $q=1$.
%
%We derive regular solutions for such skyrmions and address their incompatibility with the uniform host state.
%, derive regular solutions for asymmetric skyrmions embedded in the cone phase.
We find that---unlike the repulsive axisymmetric skyrmions and the
attractive asymmetric skyrmions respectively embedded in the PFM
state and the conical phase of chiral magnets---the asymmetric
skyrmions emerging in the TFM state of polar magnets exhibit
anisotropic inter-skyrmion potential. Depending on the relative
orientation of the two individual skyrmions, this potential can be
attractive, leading to the formation of biskyrmion or multiskyrmion
states, or repulsive.

%%%%%%%%%%%%%%%%%%%%%%%%%%%%%%%%%%%%%%%%%%%%%%%%%%%%%%%%%%%%%%%%%%%%%%%%%%%%%%%%%%%%%%%%%%%%%%%%%%%%%%%%%%%%%%%%%%%%%55
%2. Model
%%%%%%%%%%%%%%%%%%%%%%%%%%%%%%%%%%%%%%%%%%%%%%%%%%%%%%%%%%%%%%%%%%%%%%%%%%%%%%%%%%%%%%%%%%%%%%%%%%%%%%%%%%%%%%%%%%%%%%%%

Chiral solitons and modulated phases can be derived by minimizing
the energy functional of a non-centrosymmetric ferromagnet
\cite{Dz64,Bak80,thesis}:% we write  the magnetic energy density for a cubic helimagnet with uniaxial distortions along $z$-axis as
\begin{equation}
w=\sum_{i,j}(\partial_i m_j)^2 - k_u m_z^2
-\mathbf{m}\cdot\mathbf{h}+w_D\,. \label{density}
\end{equation}
Here, we use reduced values of the spatial variable,
$\mathbf{x}=\mathbf{r}/L_D$ with $L_D=A/D$ being the periodicity of
the modulated states.
$A$ is the exchange stiffness constant. The sign of the
Dzyaloshinskii constant $D$ determines the sense of rotation. %, while the periodicity is proportional to $A/D$.
$k_u=K_uA/D^2<0$ is the uniaxial anisotropy of easy-plane type,
$\mathbf{m}=[\sin\theta\cos\psi, \sin\theta\sin\psi, \cos\theta]$ is
the unity vector along the magnetization, and
$\mathbf{h}=\mathbf{H}A/D^2$ is the applied magnetic field in
reduced units.

Depending on the crystal symmetry, the DMI energy $w_D$ includes
certain combinations of Lifshitz invariants $\mathcal{L}^{(k)}_{i,j}
= m_i \partial m_j/\partial x_k  - m_j
\partial m_i/\partial x_k $ \cite{Dz64,Bogdanov89}.
Particularly, for cubic helimagnets, such as MnSi
\cite{Muehlbauer09}, FeGe \cite{Yu11}, Cu$_2$OSeO$_3$ \cite{Seki12}
and $\beta$-type Mn alloys \cite{Tokunaga15}, belonging to the
chiral $23$ (T) and 432 (O) crystallographic classes, the DMI is
reduced to the following form
\begin{equation}
w_D=\mathbf{m} \cdot (\nabla\times\mathbf{m})
\label{Bloch}
\end{equation}
and stabilizes Bloch-type modulations. %, e.g. skyrmions and spirals (helices), shown in Fig. \ref{FigBloch}.
Two types of isolated  skyrmions have been found to exist for DMI
(\ref{Bloch}): axisymmetric skyrmions within the PFM state and
asymmetric skyrmions within the conical phase, respectively
visualized in Fig. \ref{FigBloch} (a) \& (b).

In the phase diagram  of noncentrosymmetric ferromagnets with DMI
(\ref{Bloch}) and easy-plane uniaxial anisotropy, the conical phase
is the only stable modulated state \cite{Wilson2014,Butenko10,Rowland2016}.
Other states including skyrmion lattices and spirals are only
metastable solutions of the model, though can be stabilized in real
materials by several means, such as thermal fluctuations and
confined geometry.
%
%Therefore,
Axisymmetric isolated skyrmions exist as metastable excitations of
the PFM state for $h>0.5-2k_u$, while asymetric skyrmions are
present for $h<0.5-2k_u$ \cite{Wilson2014}.
%
%For a considered case of easy-plane anisotropy, the conical phase is the only thermodynamically stable state of a system.
%

\begin{figure}
\includegraphics[width=0.95\columnwidth]{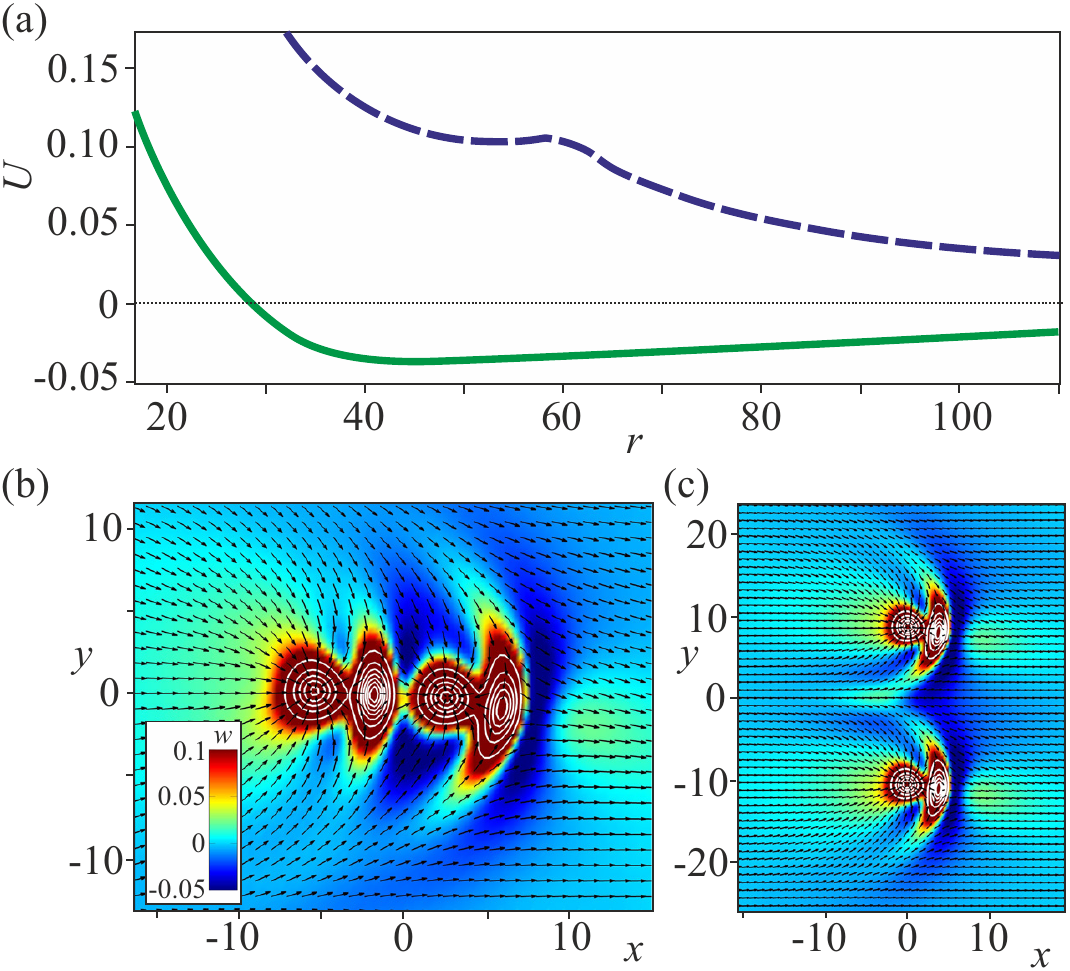}
\caption{ (color online) (a) The inter-skyrmion potential $U$ versus
the distance $r$ between the skyrmion centers. $U(r)$ was calculated
by imposing the constraint, $m_z=-1$, at the skyrmion centers \cite{thesis} and
minimizing the energy with respect to spins at all other sites
($h=0.5,\,k_u=-1.1$). Contour plots of the energy density
distribution $w(x,y)$ for (b) a bound pair of asymmetric skyrmions
in the head-to-head configuration and (c) a repulsive pair of
asymmetric skyrmions in the side-by-side configuration. Black arrows
represent the in-plane magnetization components.
The energy scale, common for panels (b) \& (c), is chosen in the
range $(-0.05,0.1)$ to highlight the inter-skyrmion regions.}
%Black arrows should be either removed from panels (b) and (c), or should be made bigger in panel (c). If you decide to erase them, please remove the corresponding sentence from the caption.
\label{cluster}
\end{figure}

The form of DMI is different for materials belonging to polar
uniaxial (C$_{nv}$) crystallographic classes,
\begin{equation}
w_D=m_x\partial_xm_z-m_z\partial_xm_x+m_y\partial_ym_z-m_z\partial_ym_y,
\label{DMI}
\end{equation}
which are the subject of the present work. This type of DMI can
stabilize skyrmions and spirals (cycloids) of N\'eel-type with the
rotation plane of the magnetization including the wave vector and
the polar axis (Fig. \ref{structure}).
As the conical phase is suppressed in this case, the phase diagram
reveals wide regions of different states with one- and
two-dimensional modulations, as discerned in Figs. 1 and 4 of the
Supplement \cite{Rowland2016,Tretiakov2016,Banerjee2014}.
The TFM and PFM states in polar magnets with easy-plane anisotropy
host two distinct types of isolated skyrmions, as shown in Fig.
\ref{structure}.

%%%%%%%%%%%%%%%%%%%%%%%%%%%%%%%%%%%%%%%%%%%%%%%%%%%%%%%%%%%%%%%%%%%%%%%%%%%%%%%%%%%%%%%%%%%%%%%%%%%%%%%%%%%%%%%%%%%%%%%%
%3. Phase diagram
%%%%%%%%%%%%%%%%%%%%%%%%%%%%%%%%%%%%%%%%%%%%%%%%%%%%%%%%%%%%%%%%%%%%%%%%%%%%%%%%%%%%%%%%%%%%%%%%%%%%%%%%%%%%%%%%%%%%%%%%

Isolated axisymmetric N\'eel skyrmions within the PFM phase are
characterized by azimuthal ($\theta$) and polar ($\psi$) angles of
the spins according to
\begin{eqnarray}
\theta = \theta (\rho), \quad \psi = \varphi. \quad
\label{skyrmion1}
\end{eqnarray}
Here the boundary conditions are $\theta (0) = \pi$, $\theta
(\infty) = 0$, while $\varphi$ and $\rho$ are cylindrical
coordinates of the spatial variable.

On the other hand, the isolated N\'eel skyrmions embedded in the TFM
phase are confined by the following in-plane boundary conditions:
\begin{eqnarray}
\theta (0) = \pi, \theta (\infty) = \theta_{TFM}=\arccos (h/2k_u). %, \, \quad \psi = \varphi, \quad
\label{skyrmion2}
\end{eqnarray}
%
%Skyrmions in the TFM phase have the antiparallel magnetization on
%the skyrmion axis ($\theta|_{\rho=0}=\pi$) !!!should be defined!!!
%and approach the uniform pattern of the TMF phase at large distances
%from the skyrmion axis:
%
%\begin{equation}
%\cos(\theta_{TFM})=h/2k_u.
%\label{boundary}
%\end{equation}
%
These boundary conditions violate the rotational symmetry, forcing the skyrmions to develop an asymmetric shape. % with axisymmetri internal structure.

The results of the numerical minimization of the energy functional
(\ref{density}) with boundary conditions (\ref{skyrmion2}) are shown
in Figs. \ref{structure} (a) \& (b).
When the canted moment of the TFM state is along the $y$ axis, the
asymmetry is clearly reflected in both $m_x$ and $m_z$. As implied
by $m_z$, such skyrmions consist of a strongly localized nearly
axisymmetric core and an asymmetric transitional region toward TFM
state.
From the left side of the depicted skyrmion, the magnetization
rotates directly from $\theta_{TFM}$ to $\theta=\pi$ in the skyrmion
center.
In contrast, at the right side, the magnetization first passes
through $\theta=0$ ($m_z=1$) and then converges back to
$\theta_{TFM}$.
This is the reason of the crescent-shaped anti-skyrmion-like region
with a positive energy density over the TFM state, shown in Figs.
\ref{cluster} (b) \& (c), which is necessary to maintain the
topological charge $q=1$ of such a highly distorted asymmetric
skyrmion.

As already mentioned, N\'eel skyrmions in insulating hosts can have
a polar dressing in addition to their magnetic pattern. The
asymmetry of N\'eel skyrmions embedded in the TFM phase is also
reflected in the spatial pattern of their electric polarization, as
visualized in Fig. \ref{structure} (c) in comparison with the polar
pattern of an axisymmetric N\'eel skyrmion in Fig. \ref{structure}
(f). The magnetically induced polarization was calculated using the
lowest-order magnetoelectric terms allowed in materials with
C$_{4v}$ or C$_{6v}$ symmetry:

\begin{eqnarray}
\mathbf{P}=\left[\alpha m_xm_z,\alpha m_ym_z,\beta m_z^2+\gamma
(m_x^2+m_y^2)\right]. \label{polarization}
\end{eqnarray}

For the polarization patterns shown in Figs. \ref{structure} (c) and
(f), we set $\alpha$=$\beta$=-$\gamma$ and used the normalization
condition, $|\mathbf{m}|$=1. While the feasibility of
electric-field-driven switching has been demonstrated already for
N\'eel skyrmions \cite{Hsu17}, the asymmetry of the polarization
pattern, characteristic to the isolated skyrmions studied here, can
be exploited to control their orientation by in-plane electric
fields.

\begin{figure}
\includegraphics[width=0.95\columnwidth]{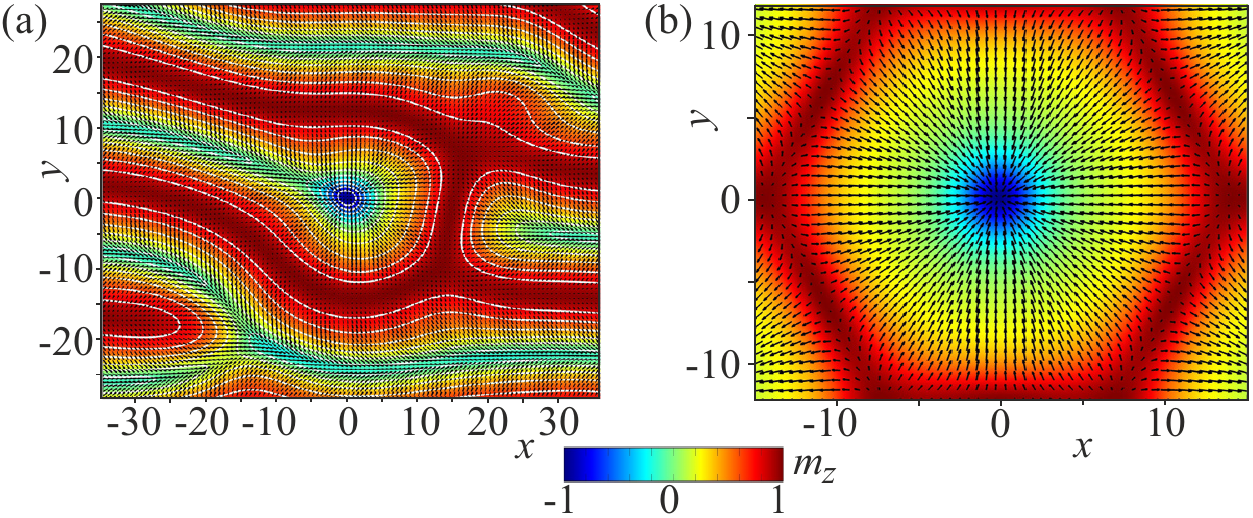}
\caption{ (color online) (a) Instability of an isolated N\'eel
skyrmion with respect to the elliptical cone state as obtained for
$h=0.5$ and $k_u=-0.5$. (b) The structure of skyrmions in the
hexagonal skyrmion lattice for $h=0.5$ and $k_u=-0.7$.} \label{PD}
\end{figure}

Asymmetric skyrmions within the TFM phase can be considered as $xy$
cross-sections of asymmetric skyrmions in the conical phases, shown
in Figs. \ref{FigBloch} (b) \&  (d) \cite{LeonovJPCM16,LeonovAPL16}.
However, the DMI term (\ref{DMI}) stabilizing N\'eel skyrmions does
not support any modulation along the $z$ axis. Thus, the asymmetry
of N\'eel skyrmions created within the TFM state is uniform along
the $z$ axis, which results in a non-trivial character of
inter-skyrmion potential, displayed in Fig. \ref{cluster} (a).
Figs. \ref{cluster} (b) \& (c) present energy density distributions
in skyrmion pairs for two mutual orientations, head-to-head and
side-by-side.
In the head-to-head configuration, skyrmions form pairs with a fixed
inter-skyrmion distance, implying the attractive nature of their
interaction, as clear from the green curve in Fig. \ref{cluster}
(a). Therefore, these skyrmions are expected to form 1D chains
running along the canted magnetization component of the TFM phase
\cite{Lin2015}.
The calculated inter-skyrmion potential for the side-by-side
configuration, the blue curve in Fig. \ref{cluster} (a), reveals the
repulsive character of skyrmion-skyrmion interaction at large
distances with a local minimum (or saddle point) at smaller
distances.
Such a behavior of the inter-skyrmion potential is related to the
positive and negative asymptotics of the energy density toward the
TFM state.
%  to the asymmetric distribution of the energy density in N\'eel skyrmions with positive and negative asymptotics toward the TFM phase
%
In general,  we argue that the inter-skyrmion potential inherently
contains a number of minima separated by saddle points (see also
Supplement for additional details).

In contrast, the asymmetric pattern of Bloch skyrmions is tightly
linked to the conical modulation of the host phase and rotates
around the $z$ axis in the same way for each individual skyrmions.
This synchronized screw-like rotation of the asymmetry for a pair of
such skyrmions leads to an overall attractive potential, by
averaging over head-to-head, side-by-side and intermediate
configurations alternating along the $z$ axis \cite{LeonovJPCM16}.
%

%The   underlies the attractive and the repulsive skyrmion-skyrmion interactions, correspondingly.
%

%

%shows a local minimum (or saddle point) of the inter-skyrmion potential, whereas the head-to-head configuration (Fig. \ref{cluster} (a)) is the global minimum.
%
The asymmetric skyrmions within the TFM state can exist only for
$k_u<-1$, i.e. require a relatively strong easy-plane anisotropy. %(region with orange shading in the phase diagram of Fig. \ref{PD} (a)).
For $k_u>-1$ isolated skyrmions undergo an instability towards the
elliptical cone state, as demonstrated in Fig. \ref{PD} (a) (see
Ref. \onlinecite{Rowland2016} and Supplement for a detailed information on
the structure of the elliptical cone and its lability region).
This instability resembles the elliptical instability of skyrmions
into spirals considered in Ref. \onlinecite{LeonovNJP16}
and allows to generalize the considered phenomenon: isolated
skyrmions tend to elongate into one-dimensional states (elliptical
cones or spirals) which have smaller energy for given control
parameters.

%This instability resembles the elliptical instability of skyrmions into spirals \cite{LeonovNJP16}.
%
%Whenever rotational symmetry is broken, isolated skyrmions tend to
%elongate into one-dimensional states (elliptical cones or spirals)
%which have smaller energy for given control
%parameters. % \cite{Karube et al., Nanoletters, CoZnMn}.
%

The TFM state turns into the PFM state with $\theta_{PFM}=0$ at the
line $h=2k_u$.
In the PFM state the rotational symmetry is recovered and isolated
skyrmions become axisymmetric with $\theta=\theta(\rho)$ and
$\psi=\phi$, as shown in Figs. \ref{structure} (d) \& (e).
%
%Cylindrical coordinates $(\rho,\phi,z)$ can again be adopted to
%these isolated skyrmions.
%
The core region and the surrounding ring have positive and negative
energy densities, respectively, implying a repulsive
skyrmion-skyrmion interaction \cite{Bogdanov94}.

N\'eel skyrmions can also form the thermodynamically stable skyrmion
lattices.
%
%having no opportunity to form a stable skyrmion lattice (due to e.g. high coercitivity as in Ref. \onlinecite{LeonovNJP16}),
%
Skyrmions within unit cells of such lattices have perfectly
hexagonal shape, as seen in Fig. \ref{PD} (b), and do not bear any
hint on the asymmetric skyrmion structure or skyrmion instability
into the elliptical cone.

%%%%%%%%%%%%%%%%%%
%discussion
%%%%%%%%%%%%%%%%%%%%%%%%%%%%%%%%%%%%%%%%%%%%%%%%%%%%

Results obtained  within the model (\ref{density}) with DMI
(\ref{DMI}) are valid for \textit{bulk} polar magnets with axial
symmetry as well as for \textit{thin films} with interface induced
DMI.
In particular, bulk polar magnetic semiconductors GaV$_4$S$_8$
\cite{Kezsmarki15} and GaV$_4$Se$_8$ \cite{Bordacs17} with the
$C_{3v}$ symmetry possess uniaxial anisotropy of easy-axis and
easy-plane type, respectively.
Since the magnitude of the effective anisotropy in these lacunar
spinels strongly varies with temperature, these material family
provides an ideal arena for the comprehensive study of anisotropic
effects on modulated magnetic states \cite{Bordacs17}.
%
%The DMI of the form  (\ref{DMI}) alternatively arises due to Rashba spin-orbit coupling (SOC) in systems that break surface mirror symmetry \cite{Tretiakov2016,Rowland2016}.
%
Skyrmions were also studied experimentally in various systems with
interface induced DMI \cite{Romming13,Romming15,Dupe16,Woo16}.
%
%Such systems typically consist of a single or double atomic layers
%of different atomic species which are deposited successively onto a
%nonmagnetic supporting crystal.
%
In these thin film structures, the rotational symmetry can also be
broken by different anisotropic environments due to lattice strains
or reconstructions in the magnetic surface layers
\cite{Hagemeister16}, as has been discussed recently for the double
atomic layers of Fe on Ir(111) \cite{Hsu16}. These structural
anisotropies also promote the formation of asymmetric skyrmions.

In conclusion, we found a new type of isolated skyrmions emerging in
tilted FM states of polar magnets with easy-plane anisotropy.
These novel solitonic states are characterized by an asymmetric
shape and an anisotropic inter-skyrmion potential.
%
%Skyrmionic states within the conical phases of cubic helimagnets have the same structure for a fixed coordinate $z$, but by additional rotation along $z$ such skyrmions average out the anisotropy inherent to considered skyrmions within the TFM state.
%
Our results are of particular interest for 2D materials like thin
films, surfaces, interfaces, where easy-plane anisotropy can coexist
with Rashba-type spin-orbit coupling, activated by the broken
surface-inversion symmetry \cite{Rowland2016,Banerjee2014}.
In order to fully explore their characteristics and functionalities,
the internal structure of these asymmetric skyrmions should be
studied experimentally, as was done for the axisymmetric individual
skyrmions within polarized FM states \cite{Romming13,Romming15}.

The authors are grateful to K. Inoue, A. Bogdanov, and Y. Togawa for
useful discussions. This work was funded by  JSPS Core-to-Core
Program, Advanced Research Networks (Japan). This work was supported
by the Hungarian Research Fund OTKA K 108918.

\section{Supplementary Information}

\begin{figure}
\includegraphics[width=0.8\columnwidth]{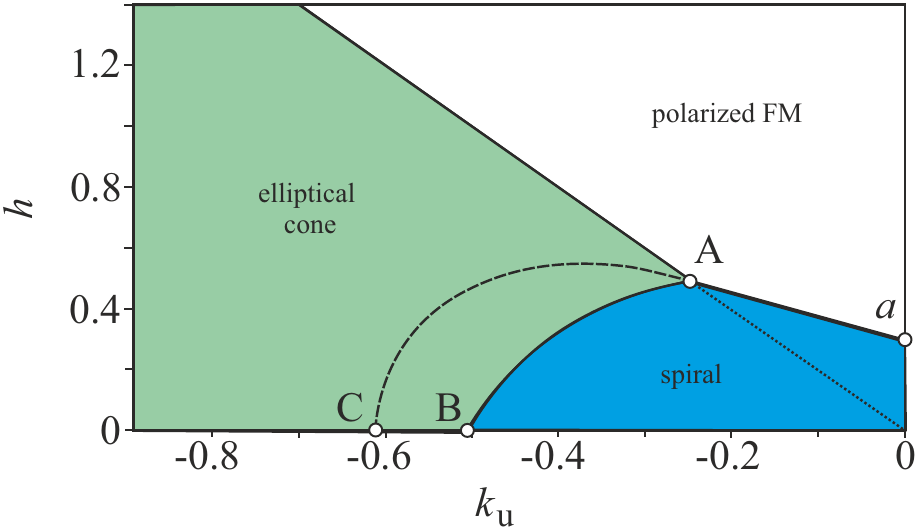}
\caption{ 
(color online) (a) Magnetic phase diagram of the solutions for model (1) with easy-plane uniaxial anisotropy. Only homogeneous and one-dimensional modulated structures are taken into account. Filled areas designate the regions of thermodynamical stability of corresponding phases: blue shading - cycloidal spiral, green shading - elliptical cone, white shading - polarized ferromagnetic state. Line $A-B$ indicates the second-order phase transition between two spiral phases. At the line $a- A - C$ the cycloidal state infinitely expands into the homogeneous state, although the line $A - C$ is never reached and represents a theoretical result  for cycloids with fixed boundary conditions $\theta(0)=0,\, \theta (p/2)=\pi$ obtained in Ref. \onlinecite{Wilson2014}. 
\label{PD}
}
\end{figure}

\begin{figure}
\includegraphics[width=0.8\columnwidth]{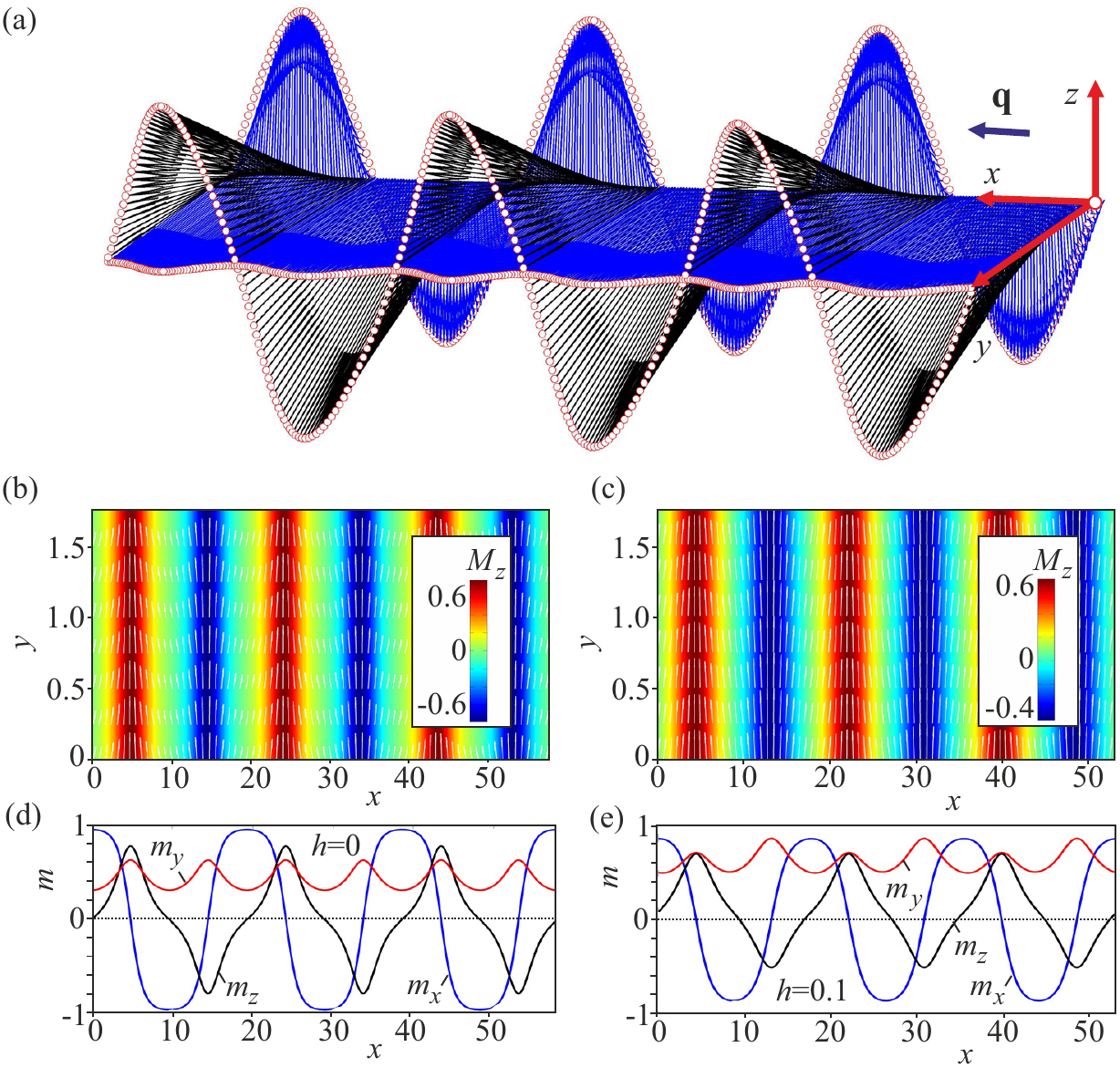}
\caption{ 
(color online)  (a) The structure of the elliptical cone state in chiral magnets with easy-plane anisotropy constructed for $k_u=-0.6,\, h=0$. The $\mathbf{q}$-vector is directed along $x$ axis. In the elliptic cone, the magnetization (shown as black arrows and additionally highlighted by red circles at their heads) traces out an elliptic cone, i.e., the cross section is an ellipse rather than a circle. The blue arrows show the corresponding projections on the $xz$ and $xy$ planes. The magnetization components $m_x(x)$ and $m_z(z)$ (blue arrows in the $xz$ plane) give rise to DMI (\ref{DMI2}); the in-plane $m_x$, $m_y$ components (blue arrows in the $xy$ plane) define the energy of the uniaxial anisotropy.  
(b), (c) Variations of the magnetization component $m_z$ along the $x$ coordinate  (color plot) together with in-plane magnetization components (white arrows) are shown for two values of the field, $h=0$ and $h=0.1$, respectively. (d) and (e) Corresponding magnetization components $m_x,\,m_y,\,m_z$ as functions of $x$.
\label{structure1}
}
\end{figure}

\subsection{Elliptical cone}

\begin{figure}
\includegraphics[width=0.8\columnwidth]{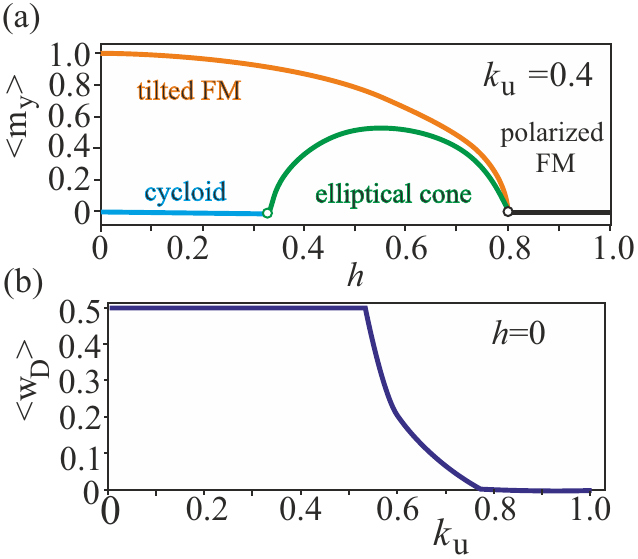}
\caption{ 
(color online)  
(a) the field-dependence of the average $m_y$-component shown in the cycloidal state, elliptical cone, and the TFM state. The color markings of the curves correspond to the colors of the stability regions at the phase diagram. $m_y=0$ in cycloids and polarized FM state; it is non-zero in TFM and elliptical cone states. (b) The rotational DMI energy is shown in dependence on the value of $k_u$ for $h=0$. The DMI energy steeply goes to $0$ around $k_u=-0.75$ although the elliptical cone as the global minimum of the system exists till $k_u=-1$.
\label{structure2}
}
\end{figure}

We analyze solutions of the model (1) starting from one-dimensional cycloids and elliptical cones (Fig. \ref{PD}). % (phase diagram in Fig. \ref{PD} (a) and Fig. \ref{structure}).
We direct their $\mathbf{q}$-vectors along $x$, and thus get the polar and azimuthal angles as functions of only one spatial coordinate: $\theta=\theta(x), \psi=\psi(x)$. 
The twisting DMI acquires the following form:
\begin{equation}
w_D=-\cos(\psi) \partial_x\theta+(1/2) \sin(2\theta)\sin(\psi)\partial_x\psi
\label{DMI2}
\end{equation}
For a cycloid, $\psi=0$ and thus $w_D=\partial_x\theta$ (the notation $\partial_x=\partial/\partial x$ is used). 
For $h=0$ the average value of DMI energy equals $1/2$ like in undistorted spirals with uniform rotation (Fig. \ref{structure2} (b)).
In the applied magnetic field, the rotational part of the cycloids is squeezed to narrow domain walls between wide domains polarized along the field. %due to the larger fraction of the magnetization pointing along the field, 
Thus, the rotational energy gradually decreases to 0. 
Usually, such a cycloid is considered to satisfy the boundary conditions: $\theta(0)=0,\,\theta(p/2)=\pi$ where $p$ is a pitch of a cycloid subject to minimization.
Thus, in the case of the easy-plane   uniaxial anisotropy  as well as in the case of easy-axis anisotropy \cite{Butenko10}, the cycloid is believed to infinitely expand its period and transform into the TFM state at the line $a-A-C$ (Fig. \ref{PD}; see also Fig. 12 in Ref. \onlinecite{Wilson2014}).

In our simulations, however, we show that the cycloidal spiral rather gives rise to the elliptical cone (Fig. \ref{structure1}) at the line $A-B$ (Fig. \ref{PD}). 
An elliptical cone (EC) was introduced in Ref. \onlinecite{Rowland2016}. 
Spins in this phase trace out a cone with an elliptical cross-section (Fig. \ref{structure1} (a) - (e)). 
The angle of this cone is defined by the corresponding angle of the tilted FM state, $\theta=\arccos(h/2k_u)$. 
Such an EC develops from a cycloidal state by the second-order phase transition and gradually increases its $m_y$-component (Fig. \ref{structure2} (a)).
At the line $h=2k_u$, EC continuously transforms into the polarized FM state (Fig. \ref{PD}).
For $k_u<-1$ the EC transforms into the tilted FM state. We note that the TFM state is isotropic in plane. 
On the contrary, the EC develops its in-plane component perpendicular to the $\mathbf{q}$-vector.
By this, the uni-directional sense of the magnetization rotation in EC is preserved, and the DMI energy of EC (\ref{DMI2}) is modified by the additional $\psi(x)$-dependence. At the lines $k_u=-1$ and $h=2k_u$ the rotational energy (\ref{DMI2}) falls to $0$ (Fig. \ref{structure2} (b)).   

\subsection{The phase diagram of states}

\begin{figure}
\includegraphics[width=0.8\columnwidth]{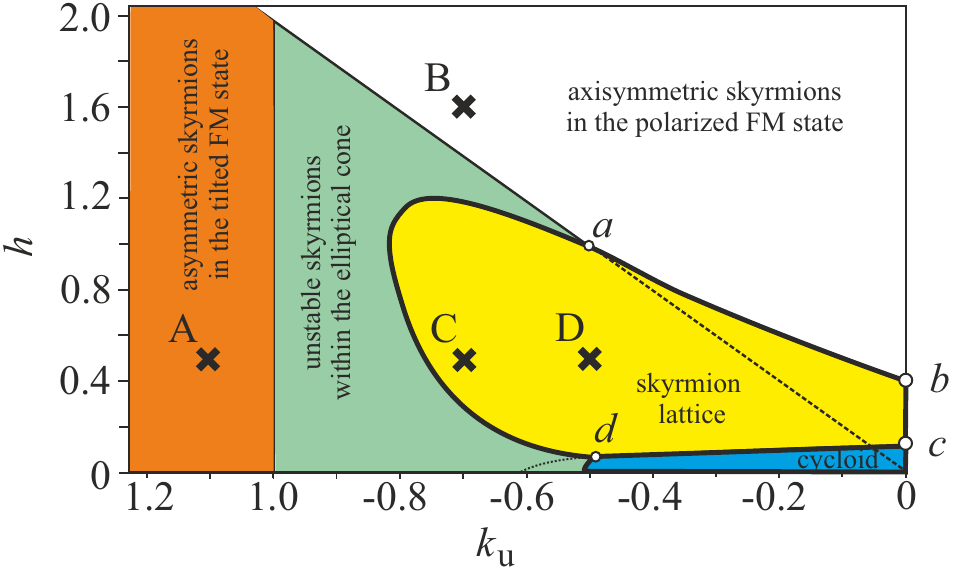}
\caption{ 
(color online) Magnetic phase diagram of the solutions for model (1) with easy-plane uniaxial
anisotropy and DMI (3).  Filled areas designate the regions of thermodynamical stability of corresponding phases: blue shading -
cycloidal spiral, green shading - elliptical cone, white shading - polarized ferromagnetic state, yellow shading - hexagonal skyrmion
lattice, orange shading - TFM state. Thick black lines indicate the first-order phase transitions between corresponding phases, thin
black lines - the second-order phase transitions.
\label{PD2}
}
\end{figure}

The complete phase diagram (Fig. \ref{PD2}) of states of the model (1) has been reproduced from Ref. \onlinecite{Rowland2016} and includes stability regions of modulated phases and regions of metastable skyrmions. 
Points A-D indicate parameters for different solutions used in the manuscript: point A -- asymmetric skyrmions within TFM phase (Fig. 2 (a) - (c)); point B -- axisymmetric skyrmions within PFM phase (Fig. 2 (d) - (f)); point C - hexagonal skyrmion lattice (Fig. 4 (b)), point D - instability of asymmetric skyrmions with respect to the elliptical cone (Fig. 4 (a)).

The phase diagrams also allows to generalize the processes of skyrmion lattice formation.
Along the line $a - b$ the skyrmion lattice appears as a result of condensation of isolated skyrmions (building blocks of the
hexagonal skyrmion lattice), as found for axisymmetric skyrmions in
the easy-axis case \cite{Bogdanov94,Butenko10,LeonovNJP16}.
Along the line $c -d$, hexagonal skyrmion lattice may appear as a
result of local cutting of the cycloid (in this sense, two merons
may be considered as nuclei of the skyrmion lattice \cite{Ezawa11}).
Along the first-order phase transition line $a - d$, however, none
of the aforementioned mechanisms is appropriate.
Presumably, domains of the skyrmion lattice and the elliptical cone
state coexist with non-trivial domain boundary between them.

\subsection{Inter-skyrmion potential}

To calculate the skyrmion-skyrmion interaction potential,  the following procedure was proved to be appropriate for axisymmetric skyrmions \cite{LeonovNC15} and asymmetric skyrmions within the conical phase \cite{LeonovJPCM16}:
the energy density (1) is minimized  with the constraint $m_z=-1$ imposed at the centers of two skyrmions as a function of the distance between skyrmion centers.
This procedure, however, faces the following difficulty when applied to asymmetric skyrmions within TFM states:
skyrmions will locally deform the surrounding TFM state to achieve the minimum of the system.
Thus, only positions of (local and global) minima  can be found precisely.
To reconstruct the interaction potential with all underlying details, one should impose a control over the in-plane component of the TFM state, i.e. violate its in-plane isotropy.

\end{document}